\documentclass{mn2e}
\usepackage{times}
\input{psfig.sty}
\newif\ifAMStwofonts

\title{The discovery of X-ray binaries in the Sculptor Dwarf
Spheroidal Galaxy}

\author[Maccarone et al.] {Thomas J. Maccarone\\ Astronomical
Institute ``Anton Pannekoek'', University of Amsterdam, Kruislaan 403,
1098 SJ, Amsterdam, The Netherlands\\ and \\ School of Physics and
Astronomy, University of Southampton, Southampton, Hampshire, SO17 1BJ, tjm@astro.soton.ac.uk
\newauthor Arunav Kundu and Stephen E. Zepf\\ Department of Physics
and Astronomy, Michigan State University, East Lansing, MI, USA
\newauthor Anthony L. Piro\\ Department of Physics,
University of California at Santa Barbara, Santa Barbara, CA, USA
\newauthor Lars Bildsten\\ Department of Physics and Kavli Institute of Theoretical Physics, University of California at Santa Barbara, Santa Barbara, CA, USA
}

\date{}

\begin{document}

\maketitle

\label{firstpage}

\def\simlt{\mathrel{\rlap{\lower 3pt\hbox{$\sim$}}
        \raise 2.0pt\hbox{$<$}}}
\def\simgt{\mathrel{\rlap{\lower 3pt\hbox{$\sim$}}
        \raise 2.0pt\hbox{$>$}}}

\input epsf

\begin{abstract}
We report the results of a deep Chandra survey of the Sculptor dwarf
spheroidal galaxy.  We find five X-ray sources with $L_X$ of at least
$6\times10^{33}$ ergs/sec with optical counterparts establishing them as
members of Sculptor.  These X-ray luminosities indicate that these
sources are X-ray binaries, as no other known class of Galactic point
sources can reach 0.5-8 keV luminosities this high.  Finding these
systems proves definitively that such objects can exist in an old
stellar population without stellar collisions.  Three of these objects
have highly evolved optical counterparts (giants or horizontal branch
stars), as do three other sources whose X-ray luminosities are in the
range which includes both quiescent low mass X-ray binaries and the
brightest magnetic cataclysmic variables.  We predict that large area
surveys of the Milky Way should also turn up large numbers of
quiescent X-ray binaries.

\end{abstract}

\begin{keywords}
X-rays:binaries -- X-rays:galaxies -- stars:binaries:close -- galaxies:individual:Sculptor dwarf spheroidal -- stellar dynamics -- stars:Population II
\end{keywords}

\section{Introduction}
It is quite difficult for field star populations to produce low mass
X-ray binaries (LMXBs), especially those with neutron star primaries.
Supernova explosions which eject more than half the mass of a system
normally leave the systems unbound.  Therefore the only ways to
produce LMXBs through binary stellar evolution are through common
envelope evolution which ejects much of the mass of the black hole's
or neutron star's progenitor before the supernova occurs (Paczynski
1976; Kalogera \& Webbink 1998), or through a finely-tuned asymmetric
velocity kick which occurs at the birth of the neutron star or black
hole (Brandt \& Podsiadlowski 1995; Kalogera 1998).  Because of these
uncertainties, and other such as the correlation between the initial
masses of the stars in a binary system, theoretical rates of X-ray
binary formation are highly uncertain.

Theoretical models suggest that the maximum rate of X-ray binary
production should occur only a few Gyrs after a burst of star
formation (e.g. White \& Ghosh 1998).  Any persistently bright LMXB
would consume the entire mass of the donor star in a few hundred
million years [i.e. the lifetime for radiatively efficient accretion
is $10^8 (10^{38} {\rm ergs/sec}/L_X$ yrs for a 1 $M_\odot$ donor
star], so the presence of bright X-ray binaries in old stellar
populations presents a challenge to theory.  With these motivations in
mind, it has been suggested that a large fraction of X-ray binaries in
the field populations of galaxies might have been produced in globular
clusters, which are known to have about 100 times as many bright LMXBs
per unit stellar mass as the field populations of the Galaxy (Katz
1975; Clark 1975).  This overabundance is likely due to dynamical
interactions which can form new X-ray binaries late in the lifetime of
a stellar population (e.g. Fabian, Pringle \& Rees 1975; Hills
1976). These dynamically formed systems would then be released into
the field either through dynamical ejections or tidal disruption of
the globular clusters in which they were born (e.g. Grindlay 1988;
White, Sarazin \& Kulkarni 2002).  Alternative possibilities are that
these systems are primarily transient accretors with very low duty
cycles, so that they can have long lifetimes as occasionally bright
accretors (Piro \& Bildsten 2002), that many are ultracompact X-ray
binaries (i.e. neutron stars or black holes accreting from white
dwarfs), which tend to form with a much longer delay from a burst of
star formation (Bildsten \& Deloye 2004), and that some LMXBs have
evolved from intermediate mass X-ray binaries, where the mass donor is
initially $1-8 M_\odot$ (e.g. Podsiadlowski et al. 2002).

The ideal sample of stars for testing the relative importance of these
different mechanisms for forming X-ray binaries would be one with no
globular clusters and a star formation history which is dominated by a
short epoch of star formation at least 8 Gyrs ago.  The system would
also need to be nearby enough that one could observe its quiescent
X-ray binaries in a reasonable exposure time.  The Sculptor dwarf
spheroidal galaxy is located at a distance of 79$\pm$4 kpc, and has
only old stellar populations, making it the only Local Group galaxy
meeting these criteria (Mateo 1998 and references within).  Recent
studies have refined the age and metallicity determinations for this
galaxy, finding that its colour-magnitude diagram is well matched by a
stellar population which formed over a 1-2 Gyr period beginning about
14 Gyrs ago (e.g. Monkiewicz et al. 1999), and with an average
metallicity of [Fe/H]=-1.4 with a dispersion of 0.2 dex
(e.g. Babusiaux, Gilmore \& Irwin 2005 and references within),
although a small tail of residual star formation continuing until
about 2 Gyrs ago cannot be ruled out with existing data (Monkiewicz et
al. 1999).

In this Letter, we report the discovery of five faint X-ray binaries
in the Sculptor dwarf spheroidal galaxy.  The presence of so many
X-ray binaries in a dwarf spheroidal galaxy which is devoid of
globular clusters indicates that binary evolution can produce a
substantial number of LMXBs without invoking stellar interactions.

\section{Data}
We have obtained 21 exposures of 6-kiloseconds each with the Chandra
X-ray observatory over the time period from 26 April 2004 to 10
January 2005.  The monitoring, rather than a single deep observation,
was performed in order to search for bright transient sources, but
none were found.  We then stacked the data to make a single image of
the data on ACIS-S3, and ran WAVDETECT, finding 74 sources in the
0.5-8.0 keV band.  Based on previous deep field measurements
(e.g. Hornschemeier et al. 2001), we estimate that about 50 of these
sources should be background AGN and that it is unlikely that more
than one is a foreground star. Given cosmic variance, the actual
number of background sources could deviate substantially from this
value.  In order to determine whether there are supersoft X-ray
sources, we have also extracted an image from 0.1-0.4 keV, but this
image contained no bright sources (i.e. with more than 10 photons)
that were not found in the 0.5-8.0 keV image.  We also ran WAVDETECT
on two other images, made with only photons from 0.5-2.0 and 2.0-8.0
keV, in order to get a crude estimate of the spectral hardnesses of
the detected sources.  None of the bright sources appear to be
strongly variable (i.e. by more than a factor of 2), while there are
too few photons in the faintest sources to search for variability.  A
more detailed discussion of variability of the bright sources will be
included in a follow-up paper.

We have compared the positions of the 9 sources with at least 100
detected counts (giving accurate positions, and making them bright
enough to rule out the possibility of a white dwarf accretor) with the
positions of bright optical stars ($V<20.5$) in the Sculptor galaxy
from Schweitzer et al. (1995).  The stars in the Schweitzer et
al. (1995) catalog are all red giants, asymptotic giant branch stars,
or horizontal branch stars.  We have also checked whether the single
brightest source has an optical counterpart in the deep optical
photometry of Hurley-Keller, Mateo \& Grebel (1999).

Below 100 counts, there also appears to be an excess compared to blank
fields, but the numbers begin to be dominated by background AGN.  The
fainter sources which are in the Sculptor galaxy could be either X-ray
binaries or bright accreting white dwarfs in the Sculptor galaxy.
While we do search for optical counterparts to these objects, we will
concentrate on the X-ray brighter sources in this Letter, and will
make a more detailed follow-up of the X-ray faint sources in future
work.

Four of the sources match the optical positions in Schweitzer et
al. within 1 arcsecond (see Table 1).  If we allow for a single
boresight correction of 0.6'' between the Chandra positions and the
positions of Schweitzer et al. (1995), then all the matches are within
0.4''.  This boresight correction is of a typical size for Chandra
(see http://cxc.harvard.edu/cal/ASPECT/celmon/).  While the 0.4''
dispersion is substantially larger than the formal uncertainties in
source positions for either the X-ray or optical catalogs, it is well
within the systematic uncertainties in source positions found by other
groups over large fields of view (see e.g. Kundu, Maccarone \& Zepf
2002; Barger et al. 2003).  The optical catalog of Schweitzer et
al. contains 529 stars within the ACIS-S3 region.  Because these are
stacked monitoring observations over a variety of roll angles, the
region is a circle of radius $4\sqrt{2}'$ rather than a square of side
8'.  The probability of a single match being a random superposition
within 1'' (to take a conservative estimate of what might be called a
match) is thus 0.0045, leading to an expected number of chance
superpositions given 9 attempts at matching of 0.04.  Alternatively,
one may consider that the next closest match after these four has a
separation of more than 2''.  It is thus unlikely that any of these
matches are chance superpositions and highly unlikely that more than
one is.  All the optical counterparts have colors and magnitudes
consistent with either the giant branch or the horizontal branch of
Sculptor, and all have proper motion estimates that indicate
probabilities of at least 96\% of being galaxy members.

The fifth match is the brightest X-ray source in our sample, which is
found at RA=1h00m13.9s, Dec=-33d44m42.5s.  While there is no strong
evidence for it to be variable in our monitoring campaign, it was not
detected in the ROSAT bright source all-sky survey, so it must either
be variable at the level of a factor of a few, or it must have a very
hard spectrum at soft X-rays. Its optical counterpart is
$R=23.68,B-R=0.95$ (D. Hurley-Keller private communication - see
Hurley-Keller et al. 1999 for a description of how the photometry was
done), giving it a luminosity and color roughly consistent with a
solar-type star in the low metallicity environment of the Sculptor
galaxy, and it appears to be near the turnoff of the main sequence in
the Sculptor galaxy.  As we do not have the full source catalogs of
Hurley-Keller et al. (1999), it is not clear what is the false match
probability for this source.  However, if it is not a real match, then
the real match is fainter than $R=23.68$, which would give a ratio of
X-ray to optical luminosity of more than 100, putting it in a region
of parameter space which is populated almost exclusively by X-ray
binaries (see e.g. Maccacaro et al. 1988).  All the secure X-ray
binaries are shown on the stacked image in Figure 1.

\begin{figure}
\psfig{width=3.5 in,file=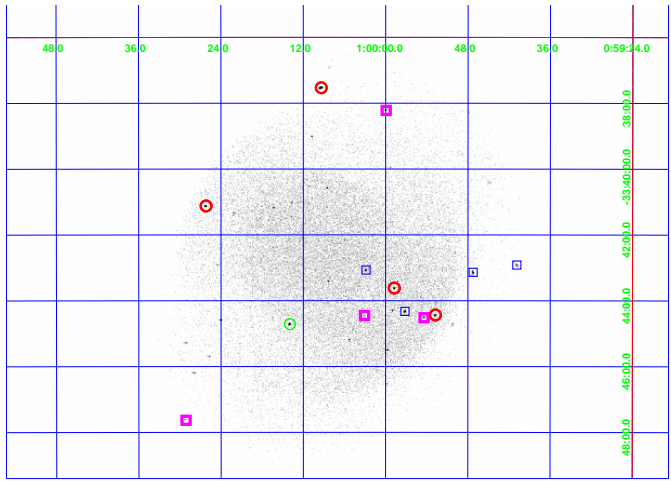}
\caption{The X-ray image of the Sculptor dwarf spheroidal galaxy.  The
thick circles show the X-ray sources with evolved counterparts from
Table 1, while the thin circle shows the source with a fainter optical
counterpart.  The less secure matches (i.e. from Table 2) are shown
with thick squares, and the bright non-matches are shown with thin
squares.}
\end{figure}

To estimate the ratio between counts/sec and ergs/sec/cm$^2$, we use
the first epoch observation of the brightest source (i.e. the one
whose optical counterpart lies close to the main sequence).  We
extract its X-ray spectrum using PSEXTRACT, and fits its spectrum in
XSPEC 11.0, and find that its count rate of 0.01 cts/sec corresponds
to a luminosity from 0.5-8.0 keV of $L_X$=5$\times10^{34}$ ergs/sec.
The spectral model which fit to this source was a power law with
$\Gamma=2.0$ and column density frozen to the Galactic value of
$2\times10^{20}$cm$^{-2}$.  We use the derived scaling between count
rate and luminosity scaling to estimate the X-ray luminosities of the
other sources shown in Table 1, although we note that most of the
sources are a bit harder than this source, so their actual
luminosities are likely to be a bit higher than those in the table.
We note that the spectra of these sources are substantially harder
than what would be expected from blackbody models with $kT=0.1-0.3$
keV.  This is not suprising, since either a black hole or a neutron
star accretor in this luminosity range is likely to be dominated by a
power law component (Jonker et al. 2004; Kong et al. 2002).  The soft,
thermal X-ray spectrum is typically only seen for neutron star
accretors below about 5$\times10^{33}$ ergs/sec, where the luminosity
can be produced by the crustal emission from the neutron star
(e.g. Brown, Bildsten \& Rutledge 1998).

We have searched the remainder of the X-ray catalog for matches as
well, and have found four more matches, although this time with a
larger matching radius of 1.5'' (since the sources have fewer counts
and hence have less well-defined X-ray positions).  One is identified
as a background galaxy, while the other three have optical colors and
proper motions which indicate that they are stars in the Sculptor
galaxy.  Two of these objects are faint enough that they could
conceivably represent the bright end of the accreting white dwarf
(i.e. CV/symbiotic star) luminosity function, rather than quiescent
X-ray binaries.  However, since these objects would be in the far tail
of the white dwarf luminosity function, but in the middle of the
luminosity distribution for quiescent LMXBs, we believe they are more
likely to have neutron star or black hole accretors than white dwarf
accretors.

One of these objects is almost as bright as the sources in Table 1,
but did not make the 100 counts cut because it is located on the
periphery of the field of view in a region that was covered for less
than half the observations.  Even searching for counterparts to all 74
X-ray sources, the expected number of false sources is still only
about 0.4, so likely all the matches (or all but one) are real.
Further optical follow-up will help determine the nature of these
sources.  The properties of these sources are listed in Table 2, with
the luminosities computed using the same photon count rate to energy
flux ratio as for the sources listed in Table 1.

\begin{table*}
\begin{tabular}{rrrcccccccc}
\hline 
RA&Dec&Opt. no.& app.mag.&abs. mag.&colour&$L_X$&Soft&Hard&HR&$t_{exp}$\\ 
\hline 
01 00 09.39&-33 37 31.9&355&19.99&0.51&0.43&6$\times10^{34}$&325&82&3.9&30\\ 
01 00 26.19&-33 41 07.5&758&20.01&0.53 &0.91&1$\times10^{34}$&108&59&1.8&90\\ 
00 59 58.68&-33 43 37.1&609&20.33&0.85&0.43&1$\times10^{34}$&189&40&4.7&126\\ 
00 59 52.75&-33 44 26.1&226&19.35&-0.13&0.47&6$\times10^{33}$&73&49&1.5&126\\ 
01 00 13.84&-33 44 43.3&NA&23.68&4.18&0.95&9$\times10^{34}$ &804&212&3.8&108\\ 
\hline
\end{tabular}
\caption{The five strong X-ray binary candidates.  The columns are RA
and Declination (for the optical counterparts), Optical catalog number
from Schweitzer et al. (1995), the apparent and absolute magnitudes
[$V$ for the sources from Schweitzer et al. (1995), $R$ for the other
source], the color [$B-V$ for the sources from Schweitzer et
al. (1995), $B-R$ for the remaining source], and the X-ray luminosity
in ergs/sec, the number of counts in the soft (0.5-2.0 keV) and hard
(2.0-8.0 keV) bands, the ratio of soft to hard counts, and the
exposure time including the source's location, in kiloseconds.}
\end{table*}

\begin{table*}
\begin{tabular}{rrrcccccccc}
\hline
RA&Dec&Opt. no.& V&$M_V$ &B-V&$L_X$&Soft&Hard&HR&$t_{exp}$\\
\hline
01 00 03.03&-33 44 26.9&863&20.28&0.80&0.33& 3$\times10^{32}$&ND&ND&**&126\\
00 59 59.89&-33 38 12.2&333&20.22&0.74&0.31& 7$\times10^{33}$&47&13&3.6&42\\
00 59 54.21&-33 44 29.7&224&19.76&0.28&0.88& 2$\times10^{33}$&32&12&2.7&126\\
01 00 29.22&-33 47 37.1&1263&19.57&0.09&1.03& BG galaxy&&&&24\\
\hline
\end{tabular}
\caption{The four fainter sources with matches from the catalog of
Schweitzer at al. (1995).  The columns are RA and Declination (for the
optical counterparts), Optical catalog number, the offsets in RA and
declination (measured as Chandra position minus optical position), the
$V$ magnitude, absolute $V$ magnitude, $B-V$ color, the X-ray
luminosity in ergs/sec where the distance is known, the number of
counts in the soft (0.5-2.0 keV) and hard (2.0-8.0 keV) bands, the
ratio of soft to hard counts, and the exposure time including the
source's location, in kiloseconds.  The last of these sources is the
background galaxy.  The first source was detected only in the
full-band image, and not in the hard or soft images.}
\end{table*}

\section{Discussion}
\subsection{The nature of the matches}
First, let us convince ourselves that these sources, are in fact,
X-ray binaries.  Given that the optical counterparts fit well on the
red giant branch and horizontal branch of the Sculptor dwarf
spheroidal galaxy, and that they have proper motions which strongly
imply that they are galaxy members, they are extremely unlikely to be
background AGN.  The four systems listed in Table 1 all have X-ray
luminosities greater than $5\times10^{33}$ ergs/sec in the 0.5-8.0 keV
range.  Nearly all cataclysmic variables (see e.g. Verbunt et
al. 1997; Heinke et al. 2005) and symbiotic stars (e.g. Anderson et
al. 1981; Galloway \ Sokoloski 2004) have X-ray luminosities below
this level in the 0.5-8.0 keV band.  Non-magnetic CVs emit hard X-rays
only when they have optically thin boundary layers on the white dwarf
surface, and at high accretion rates, these boundary layers become
optically thick (e.g. Pandel et al. 2005 and references within).  Thus
the highest hard X-ray luminosities are typically found in magnetic
CVs, where much of the accretion is channelled down the magnetic poles
and hence radiates free-free emission (see King 1983 and references
within).  Suleimanov et al. (2005) quote luminosities above $10^{34}$
ergs/sec for V1223 Sgr and GK Per, but they have made large absorption
corrections and used a larger bandpass than we have.  In order to make
a consistent comparison, we needed to revise the estimates of
Suleimanov et al. (2005), and find that if in Sculptor, we
would have estimated luminosities of $6\times10^{33}$ ergs/sec.  Their
findings do indicate that perhaps the faintest source in Table 1 is
drawn from the very bright end of the magnetic CV distribution, but
our other matches are comfortably above this level.  We note also that
some symbiotic stars reach soft X-ray luminosities well in excess of
$5\times10^{33}$ ergs/sec, but these sources have very soft spectra
peaked in the far ultraviolet, and would (1) produce very few, if any,
counts in the 0.5-8 keV range, and (2) would be very strong emitters
in the 0.1-0.4 keV range.  Since we have searched for bright sources
in this lower energy range and have failed to find any that were not
also quite bright from 0.5-8.0 keV, we can rule out the existence of
such sources.  Our new detections have X-ray luminosities a bit higher
than those of most quiescent LMXBs, although since these systems are
likely to be wide binaries on the basis of their evolved optical
counterparts, this is perhaps not a surprise given the correlation
between quiescent luminosities of X-ray binaries and their orbital
periods (see e.g. Garcia et al. 2001).

It has been suggested that the field populations of elliptical
galaxies are low duty cycle transients, with red giant donors (Piro \&
Bildsten 2002).  Obviously the finding that several of these systems
have red giant donors provides support for this hypothesis.  The
number of bright (i.e. $L_X>10^{37}$ ergs/sec) LMXBs per unit stellar
mass seems to be roughly constant across a sample of giant elliptical
galaxies (Gilfanov 2004), with a typical value of about one per
$10^8-10^9 M_\odot$ of stars.  We would then expect about $\sim$ .02
such systems in Sculptor, given its mass of about 2$\times10^6
M_\odot$ (Mateo 1998).  Given that the duty cycles of these transients
are likely to be less than about 1/200 in the scenario of Piro \&
Bildsten (2002), one would expect that the number of quiescent
transient X-ray binaries in the Sculptor galaxy would be at least a
few, and possibly much higher if the duty cycles were even smaller
than 1/200.  Our discovery of one clear case of an X-ray binary with a
red giant counterpart and a few more candidates is thus in reasonable
agreement with the picture suggested by Piro \& Bildsten (2002).

Several matches appear to be associated with horizontal branch, rather
than giant branch stars (i.e. those with $V>20$ and $B-V<0.5$).  Given
that the photometric errors at limit of the photometry used are about
0.1 magnitudes (Schweitzer et al. 1995), some of these could be giant
stars shifted onto the horizontal branches by measurement
uncertainties, but Washington photometry (Westfall et al. 2005) and
infrared photometry (Babusiaux et al. 2005) also find that source 609
from Schweizter et al (1995) seems to be a horizontal branch star.  It
therefore seems that horizontal branch stars also may be the donor
stars in X-ray binaries.  This might be indicating that mass loss from
the accretion process has affected the structures of these stars
substantially, yielding a donor star similar to that in Cygnus X-2,
which appears in a colour-magnitude diagram at a location near the
horizontal branch, despite probably having had a very different
evolutionary history (see e.g. Podsialowski \& Rappaport 2000; Kolb et
al. 2000).

It is also worth noting that if the accretors are black holes, optical
light from the accretion flow could contaminate the stellar light, as
it does in V404 Cygni (Zurita et al. 2004); our newly detected systems
are about 10 times as bright in X-rays and about 3 times as bright in
the optical as V404 Cygni.  Optical contamination by the accretion
flow's light should have a smaller impact if the accretors are neutron
stars, since the optical accretion emission from neutron stars is
substantially fainter than for black holes at the same X-ray
luminosity (see e.g. Jain 2001).

\subsection{Are there more quiescent X-ray binaries in Sculptor?}
It is likely that there are actually many more quiescent X-ray
binaries in the Sculptor galaxy than the five strong matches we have
presented here and the three additional tenative matches.  Our
detection limits in this data set are at a luminosity of a little bit
above $10^{32}$ ergs/sec, while quiescent LMXBs have been seen to be
as faint as $2\times10^{30}$ ergs/sec, with most of the faintest
sources having black hole accretors (Garcia et al. 2001).  Therefore,
some quiescent X-ray binaries could be below our sensitivity limits,
although most neutron star accretors are brighter than these limits in
quiescence, and there do seem to be higher quiescent luminosities for
long period transients, even for black hole accretors.  There may also
be other quiescent X-ray binaries which are detected in our
observations, but which have faint optical counterparts.  These could
be, for example, ultracompact X-ray binaries, which have also been
suggested to be systems which should be present in old stellar
populations (Bildsten \& Deloye 2004), or even X-ray binaries with
lower main sequence donors which simply took longer than the typical
few Gyrs (White \& Ghosh 1998) to come into contact.  We plan to
investigate these possibilities in future work with deeper optical
photometry and with spectroscopic follow-ups.

\subsection{Implications for X-ray binaries in other stellar populations}
Scaling from globular clusters, it is suprising to see so many X-ray
binaries in the Sculptor dwarf spheroidal galaxy.  Globular clusters
are generally thought to be enhanced in X-ray binary production by a
factor of about 100 compared to field stars (Katz 1975; Clark 1975).
Searches for quiescent X-ray binaries in the Milky Way's globular
clusters have focused not on a representative sample of clusters, but
on the clusters where the highest rates of dynamical interactions are
expected.  Nonetheless, as an upper limit to the enhancement rate of
the production of quiescent X-ray binaries in globular clusters, we
can compare NGC~6440, the cluster with the highest stellar collision
rate, with Sculptor.  NGC~6440 has a mass of about 1/9 that of the
Sculptor dwarf spheroidal galaxy (Harris 1996; Mateo 1998).  Using a
fainter $L_X$ limit than we have used in this paper, Heinke et
al. (2003) found 8 quiescent LMXBs in NGC~6440, so the number density
of X-ray binaries per unit stellar mass in NGC~6440 is enhanced by, at
most, a factor of 10 relative to that in Sculptor.  This is likely
related to the evolved donor stars for the X-ray binaries in Sculptor,
and hence their long orbital periods.  Such long period binaries are
susceptible both to ionization in the densest parts of a globular
cluster where their binding energies may be less than the kinetic
energies of individual stars (Heggie 1975), and to having
eccentricities induced by weak fly-by interactions, which will then
speed up their evolution (Hut \& Paczynski 1984; Rasio \& Heggie 1995;
Heggie \& Rasio 1996).  Since globular clusters contain about 100
times as many bright X-ray binaries per unit stellar mass than field
populations, but no more than 10 times as many total X-ray binaries
per unit stellar mass, it appears that the increased duty cycles of GC
LMXBs is more important than their increased number density for
explaining why there are so many more bright GC LMXBs per unit stellar
mass than field LMXBs.

Finally, we consider the implications of these results for the Milky
Way's bulge's X-ray binary population.  The bulge of the Milky Way
contains about $5\times10^2$ times as many stars as the Sculptor dwarf
spheroidal galaxy, meaning that it should have about 2,500 X-ray
binaries if the number of binaries scales linearly with mass.  In
fact, since X-ray binary production is enhanced in metal rich systems
(at least for globular clusters - see Kundu, Maccarone \& Zepf 2003
for evidence of this effect and Maccarone, Kundu \& Zepf 2004 for a
discussion of a binary evolution model for this effect which should
work equally well in field populations as in globular clusters) the
Milky Way might contain a few times more than this number of X-ray
binaries.  Our estimate is comparable to numbers from theoretical
predictions (Iben et al. 1997; Belczynski \& Taam 2004).

Finally, we note that with only about 500 optical stars in the
Schweitzer et al. (1995) catalog, at least $\approx$1\% of the giant
branch/horizontal branch stars in Sculptor are in systems with
accreting neutron stars or black holes.  Based on this finding, one
would expect that a large number of similar systems are likely to be
found in surveys that include hundreds of red giant stars.  This
hypothesis should be testable as the results from the ChaMPlane survey
(Grindlay et al. 2003) begin to come in, or in a large area survey of
the Galactic Bulge.

\section{Summary}
We find at least 5 X-ray binaries with likely neutron star or black
hole accretors in the Sculptor dwarf spheroidal galaxy.  Four of these
have evolved optical counterparts, providing support for the
hypothesis of Piro \& Bildsten (2002) that a large fraction of the
non-globular cluster X-ray sources in elliptical galaxies are wide
binary systems with red giant donors which have low duty cycles. 

\section{Acknowledgments}
We are grateful to Carine Babusiaux, Kyle Cudworth, Peter Jonker,
Christian Knigge, Mike Muno, Eric Pfahl, Andrea Schweitzer, Jeno
Sokoloski, Kyle Westfall and Rudy Wijnands for useful discussions.
We thank Denise Hurley-Keller, Eva Grebel and Mario Mateo for the
optical photometric measurement of the brightest X-ray source.  AK and
SEZ acknowledge support from NASA grants SAO G04-5091X and LTSA
NAG-12975.  LB acknowledges support from the NSF via grant PHY99-07949

\end{document}